\documentclass[article]{IEEEtran}
\IEEEoverridecommandlockouts
\usepackage{cite}
\usepackage{balance}
\usepackage{adjustbox}
\usepackage{amsmath,amssymb,amsfonts}
\usepackage{multicol}
\usepackage{algorithmic}
\usepackage{graphicx}
\usepackage{textcomp}
\usepackage{xcolor}
\usepackage{caption}
\usepackage{subcaption}
\usepackage{tabularx}
\usepackage{booktabs}
\usepackage{booktabs}
\usepackage{multirow}

\def\BibTeX{{\rm B\kern-.05em{\sc i\kern-.025em b}\kern-.08em
    T\kern-.1667em\lower.7ex\hbox{E}\kern-.125emX}}
\begin{document}

\title{ Brain Age Group Classification Based on Resting State Functional Connectivity Metrics  \\}

\author{Prerna Singh$^1*$, Kuldeep Singh Yadav$^2$, Lalan Kumar$^{1,2,3}$, and Tapan Kumar Gandhi$^{1,2,3}$\\\vspace{1mm}
    $^1$Bharti School of Telecommunication Technology and Management, Indian Institute of Technology Delhi, India\\
    $^2$Department of Electrical Engineering, Indian Institute of Technology Delhi, India\\
    $^3$Yardi School of Artificial Intelligence, Indian Institute of Technology Delhi, India\\
    *corresponding author: bsz208534@iitd.ac.in
}

\maketitle
\begin{abstract}
This study investigated age-related changes in functional connectivity using resting-state fMRI and explored the efficacy of traditional deep learning for classifying brain developmental stages (BDS). Functional connectivity was assessed using Seed-Based Phase Synchronization (SBPS) and Pearson correlation across 160 ROIs. Clustering was performed using t-SNE, and network topology was analyzed through graph-theoretic metrics. 
Adaptive learning was implemented to classify the age group by extracting bottleneck features through mobileNetV2. These deep features were embedded and classified using Random Forest and PCA. Results showed a shift in phase synchronization patterns from sensory-driven networks in youth to more distributed networks with aging. t-SNE revealed that SBPS provided the most distinct clustering of BDS. Global efficiency and participation coefficient followed an inverted U-shaped trajectory, while clustering coefficient and modularity exhibited a U-shaped pattern. MobileNet outperformed other models, achieving the highest classification accuracy for BDS. Aging was associated with reduced global integration and increased local connectivity, indicating functional network reorganization. While this study focused solely on functional connectivity from resting-state fMRI and a limited set of connectivity features, deep learning demonstrated superior classification performance, highlighting its potential for characterizing age-related brain changes.
\end{abstract}
\section{Introduction}
The global demographic shift towards an aging population has intensified research into cognitive decline and its underlying neural mechanisms. Aging is associated with progressive brain structure and function alterations, impacting cognitive domains such as memory, attention, and executive control \cite{christensen2009ageing,whalley2004cognitive}. Structural neuroimaging studies using MRI have extensively documented age-related changes, including cortical atrophy, white matter hyperintensities, and reduced grey matter volume \cite{brant1985basic}. However, these structural markers do not fully capture the functional alterations in brain network interactions that underlie cognitive aging.

Aging is a key risk factor for cognitive decline and a potential indicator of progression to neurodegenerative diseases. It can impact specific brain regions, influencing neural dynamics by modifying their activity patterns, affecting directly connected regions, altering interactions across the brain, and ultimately reshaping the entire brain network \cite{lin2016predicting}. Brain aging is a gradual process involving changes in neural function, cognition, and brain network organization \cite{lin2016predicting}. However, this process varies significantly across individuals, with some experiencing "successful aging" and others showing "accelerated aging," which is often associated with cognitive impairments and neurodegenerative diseases such as Alzheimer’s disease (AD) \cite{ritchie2001classification, swainson2001early}. Several studies suggest that neurodegenerative conditions accelerate brain aging, with observable changes occurring years before clinical symptoms appear \cite{thal2004neurodegeneration, sheng2024hybrid}. For example, AD patients exhibit an estimated 10-year acceleration in brain aging \cite{gaser2013brainage, he2024spatiotemporal}, while conditions such as noninsulin-dependent diabetes mellitus (+4.6 years) \cite{franke2013advanced}, hypertension (+4.1 years) and schizophrenia (+5.5 years) \cite{koutsouleris2014accelerated} also show deviations from typical aging trajectories. These deviations highlight the potential role of brain age as an early biomarker for neurological and psychiatric disorders.

Resting-state functional connectivity (rs-FC) has emerged as a powerful tool for examining large-scale brain network organization and its disruptions with age \cite{grady2012cognitive,he2024spatiotemporal}. Functional MRI (fMRI) studies have shown that aging is associated with reduced connectivity within the default mode network (DMN) and altered connectivity patterns in other brain networks \cite{biswal1995functional, geerligs2015brain}. In parallel, phase synchronization—reflecting the temporal coordination of neural oscillations—has gained attention as a critical marker of age-related changes in functional brain networks. Studies using phase-based connectivity measures, such as phase-locking value (PLV) and weighted phase-lag index (wPLI), have revealed a decline in long-range synchronization and an increase in local connectivity with aging, suggesting a shift from global to more fragmented network organization \cite{dennis2014functional, stam2007phase, sun2012inferring, vecchio2014human}. These changes may contribute to reduced neural efficiency and cognitive deficits commonly observed in older adults.

Recent advances in computational neuroscience have enabled the integration of deep learning models with functional connectivity metrics to improve the prediction of brain aging and neurodegenerative risk \cite{cole2017predicting,li2018brain}. Machine learning approaches leveraging graph-based and time-frequency domain features have been increasingly applied to characterize aging-related disruptions in connectivity patterns \cite{dosenbach2010prediction}. Despite these advancements, most studies still rely on coarse-grained connectivity measures that may overlook fine-scale temporal dynamics crucial for understanding cognitive decline\cite{li2018brain}.

This study aims to bridge this gap by leveraging deep learning techniques to analyze functional connectivity patterns derived from rs-FC and phase synchronization measures. It introduces an adaptive learning approach to extract and learn more prominent features for the limited samples. Various pre-designed and customized deep-learning networks were explored. Dimension reduction and traditional classification approaches were employed to mitigate the overfitting/underfitting issue during training. By integrating network neuroscience with adaptive learning, we seek to identify novel biomarkers of cognitive aging, contributing to a more comprehensive understanding of healthy and pathological aging trajectories. Our approach will enhance the predictive modeling of age-related functional brain alterations and provide insights into their implications for neurocognitive health.

The paper is structured as follows: Section II describes the materials and methods, Section III presents the results and discussion, and Section IV provides the conclusion..

\section{Materials and Methods}
Figure 1 provides a visual representation of the study's conceptual framework, outlining the key processes, methodologies, and relationships between different components of the analysis.

\begin{figure*}[t]
\centering
\includegraphics[width =1\linewidth, trim=10 70 10 82,clip]{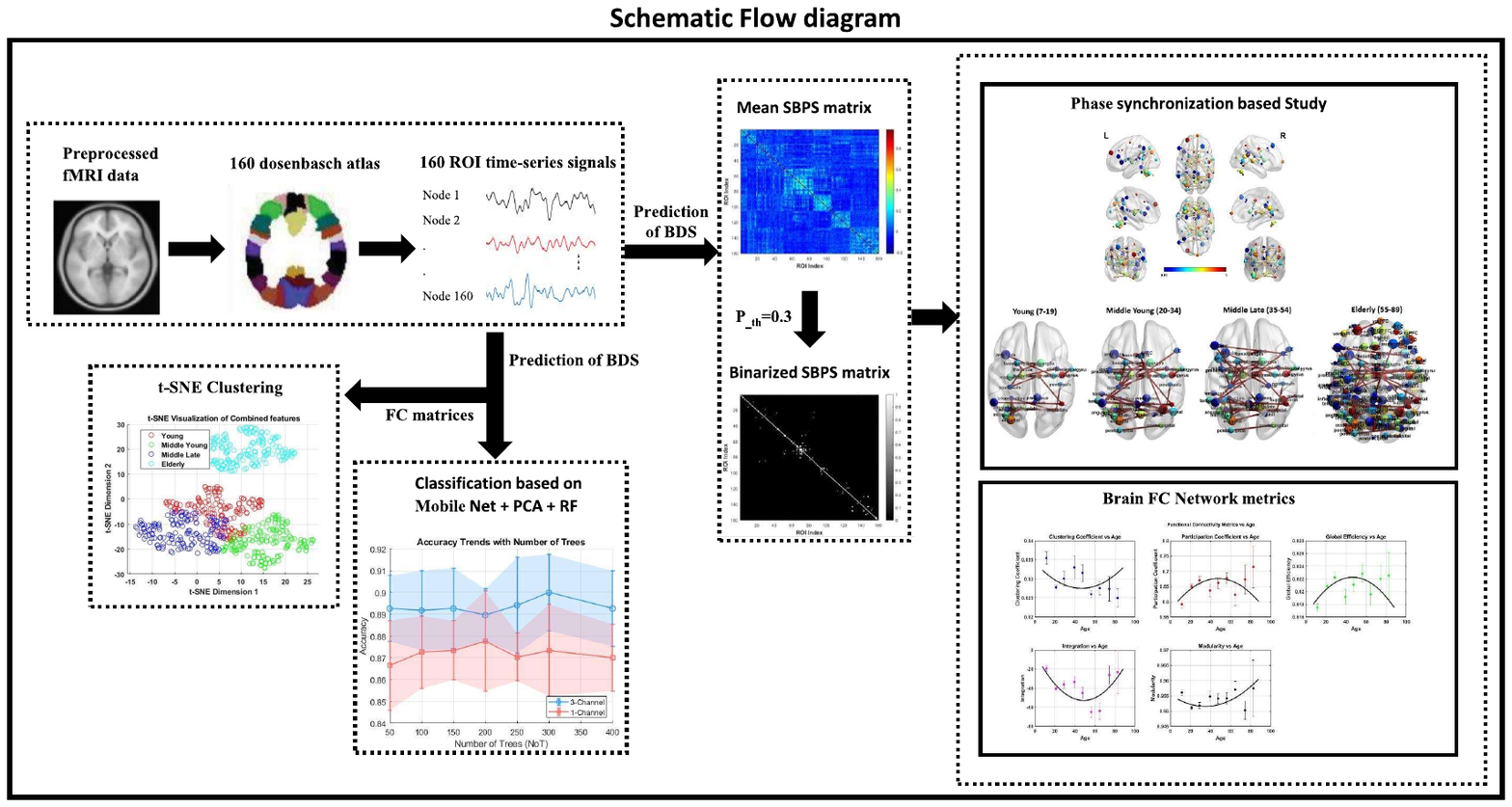}
\caption{Flow chart of the conceptual framework of the study.}
\label{figure:block diagram1}
\end{figure*}

\subsection{Dataset}
The resting-state fMRI data for this study were obtained from the Human Connectome Project (HCP) \cite{van2012human}. The primary objective of HCP was to collect data from a diverse cohort across various age groups and generate an in vivo functional connectivity map for research purposes. This publicly available dataset includes rs-fMRI scans from 1,096 healthy participants aged 7 to 89 years. Data acquisition was performed using a 3.0 Tesla Magnetom Tim Trio scanner with the following parameters: repetition time (TR) of 2530 ms, echo time (TE) of 30 ms, field of view (FOV) of 240 × 240 mm², acquisition matrix of 64 × 64, voxel size of 3 × 3 × 4 mm, and 34 slices covering the entire brain. Each scan lasted 500 seconds per subject, with participants instructed to remain awake to ensure accurate 3D functional imaging. Samples were retained after applying head motion correction and other preprocessing criteria.

Following the necessary preprocessing steps, the blood-oxygen-level-dependent (BOLD) signal time series was extracted from each voxel of the functional images \cite{di2013calibrating}. Using Dosenbach’s template \cite{dosenbach2010prediction}, 160 regions of interest (ROIs) were predefined, and the mean time series within a 5 mm spherical radius was computed. Finally, a time-series data matrix was generated for each subject, comprising 160 brain nodes across different timestamps, which was subsequently utilized in the study.

\subsection{Prediction of Brain Developmental Stages}
Building on our previous study \cite{singh2023reorganization}, which predicted brain developmental stages using resting-state fMRI data and sample entropy (SampEn) features from 160 Dosenbach ROIs, we further investigated brain network differences across these developmental stages using additional functional connectivity metrics. In our previous study,
SampEn, which quantifies signal complexity, is extracted for 1,096 subjects and combined with age information to create a dataset. K-means clustering, using the elbow criterion, identifies four developmental stages: Young (7-19 years), Middle Young (20-34 years), Middle Late (35-53 years), and Elderly (54-89 years) \cite{singh2023reorganization}. These four brain developmental stages resting state fMRI data have been used further for functional connectivity studies.


\subsection{Feature Engineering}
The accurate classification of age groups based on resting-state fMRI (rs-fMRI) requires extracting meaningful and discriminative features from the functional connectivity patterns of the brain. 
This study considered three key feature representations, i.e., a) the raw functional time-series data, b) Seed-based Phase Synchronization (SBPS), and c) the Z-transformed connectivity matrix. Each feature representation captures different aspects of functional connectivity, providing complementary information for classification. In addition, a channel-wise fusion is also incorporated to employ more prominent features.

\subsubsection{Raw Functional Time-Series Representation}
Given an rs-fMRI signal $S_i(t)$ from a region of interest (ROI) $i$, where $t$ represents the time points, the raw feature matrix $X \in \mathbb{R}^{N \times T}$ is constructed as:
\begin{equation}
    X = [S_1(t), S_2(t), \dots, S_N(t)],
\end{equation}
where $N$ denotes the number of ROIs, and $T$ is the number of time points. This representation retains the full spatiotemporal information and serves as a baseline for comparison.

\subsubsection{Seed-Based Phase Synchronization (SBPS)}
Phase synchronization is a crucial measure of functional connectivity, capturing dynamic relationships between brain regions. The analytic signal $\psi_i(t)$ of an rs-fMRI time series $S_i(t)$ can be derived using the Hilbert Transform:
\begin{equation}
    \psi_i(t) = S_i(t) + j H[S_i(t)],
\end{equation}
where $H[\cdot]$ is the Hilbert Transform. The instantaneous phase is then computed as:
\begin{equation}
    \phi_i(t) = \arg(\psi_i(t)).
\end{equation}
The phase synchronization between two ROIs, $i$ and $j$, is given by:
\begin{equation}
    \gamma_{i,j} = \left\langle \cos(\phi_i(t) - \phi_j(t)) \right\rangle_t,
\end{equation}
where $\langle \cdot \rangle_t$ denotes temporal averaging. The resulting SBPS matrix $\Gamma \in \mathbb{R}^{N \times N}$ quantifies synchronization-based functional connectivity.

\subsubsection{Z-Transformed Functional Connectivity Matrix}
Pearson correlation is widely used to construct functional connectivity matrices. The correlation coefficient between ROIs $i$ and $j$ is defined as:
\begin{equation}
    r_{i,j} = \frac{\sum_{t=1}^{T} (S_i(t) - \mu_i)(S_j(t) - \mu_j)}{(T-1)\sigma_i \sigma_j},
\end{equation}
where $\mu_i$ and $\sigma_i$ are the mean and standard deviation of $S_i(t)$. To achieve normality and stabilize variance, Fisher's Z-transformation is applied:
\begin{equation}
    Z_{i,j} = \frac{1}{2} \ln \left( \frac{1 + r_{i,j}}{1 - r_{i,j}} \right).
\end{equation}
The transformed matrix $Z \in \mathbb{R}^{N \times N}$ enhances the robustness of connectivity features by mapping them to a Gaussian-like distribution.

\subsubsection{Feature Fusion via Multi-Channel Representation (FMCR)}
The three extracted features represent different perspectives of brain connectivity, and their fusion can enhance classification performance. Instead of using them independently, we adopt a multi-channel representation where each feature type forms a distinct channel in a combined input tensor:
\begin{equation}
    F = \text{concat}(X, \Gamma, Z),
\end{equation}
where $F \in \mathbb{R}^{N \times N \times 3}$. This approach allows CNNs to exploit spatial correlations across different connectivity representations, improving feature learning. The advantage of this fusion lies in leveraging complementary information, reducing redundancy, and capturing both direct and indirect functional interactions in the brain.

Integrating these diverse functional connectivity descriptors into a unified multi-channel representation provides a more comprehensive and informative input to deep learning models, enhancing the predictive accuracy for age group classification.

\subsection{Deep Network Design and Analysis} 

Driven by the advancements in deep learning from the ImageNet Large Scale Visual Recognition Challenge (ILSVRC) \cite{ILCVRchallange}, researchers have significantly improved feature extraction and classification in various pattern recognition tasks by designing robust and lightweight deep convolutional neural networks, such as ResNet-50 \cite{ResNet}, SqueezeNet \cite{squeezenet}, MobileNetV2 \cite{mobilenetv2}, and EfficientNet-B0 \cite{efficientnet}. Some of these architectures are particularly well-suited for real-time deployment due to their efficiency and computational scalability. Table \ref{tab:imagenet_comparison} shows the performance and parametric evaluation of the ImageNet dataset \cite{ILCVRchallange}. 

\begin{table}[hbt]
    \centering
    \renewcommand{\arraystretch}{1.2} 
    \setlength{\tabcolsep}{5pt} 
    \caption{Performance comparison of deep networks on ImageNet dataset}
    \resizebox{\columnwidth}{!}{%
    \begin{tabular}{lccccc}
        \toprule
        \multirow{2}{*}{\textbf{Model}} & \multicolumn{2}{c}{\textbf{Accuracy (\%)}} & \multirow{2}{*}{\textbf{Params (M)}} & \multirow{2}{*}{\textbf{Size (MB)}} & \multirow{2}{*}{\textbf{Infer. Time (ms)}} \\
        \cmidrule(lr){2-3}
        & \textbf{Top-1} & \textbf{Top-5} & & & \\
        \midrule
        AlexNet \cite{Alexnet}         & 57.2  & 80.3  & 61.0   & 233  & 8.9  \\
        VGG-16 \cite{vgg16}         & 71.5  & 89.8  & 138.4  & 528  & 8.0  \\
        VGG-19 \cite{vgg16}         & 74.0  & 91.0  & 143.7  & 549  & 9.2  \\
        ResNet-50 \cite{ResNet}     & 76.6  & 93.1  & 25.6   & 98   & 6.0  \\
        ResNet-101 \cite{ResNet}    & 78.3  & 94.1  & 44.6   & 171  & 8.1  \\
        SqueezeNet \cite{squeezenet}    & 57.5  & 80.3  & 1.24   & 4.9  & 3.2  \\
        MobileNetV2 \cite{mobilenetv2}   & 72.0  & 91.0  & 3.4    & 14   & 4.0  \\
        EfficientNet-B0 \cite{efficientnet} & 77.1  & 93.3  & 5.3    & 20   & 5.6  \\
        DarkNet-53 \cite{darknet}     & 77.2  & 93.8  & 41.6   & 155  & 7.5  \\
        \bottomrule
    \end{tabular}%
    }
    \label{tab:imagenet_comparison}
\end{table}

Building on the strengths of state-of-the-art deep learning models, this study explored the potential of leveraging rs-fMRI data for age group classification. Specifically, the pre-trained MobileNetV2 network was extensively utilized due to its robustness and computational efficiency. This network was customized for the input $160\times160$. This network was trained and fine-tuned by leveraging transfer learning for the rs-fMRI data. This network achieved significantly high training performance. However, validating this network on the ageing dataset revealed significant overfitting, primarily due to the limited number of samples per class. 
 

Less complex architecture, fine-tuning techniques, and feature-reduction strategies were employed to mitigate this issue. This study introduced and explored customized CNN tailored for EEG-based age group classification, i.e., Shallow-CNN and DeepReg-CNN.
These models were developed to balance computational efficiency and model complexity while ensuring effective spatial feature extraction. Shallow-CNN is a lightweight architecture with minimal depth, making it suitable for computationally constrained environments. In contrast, DeepReg-CNN incorporates a deeper architecture with advanced regularization strategies, enhancing feature learning capacity and mitigating overfitting.

\subsubsection{Shallow-CNN}
The Shallow-CNN model was designed to efficiently extract spatial features while maintaining low computational complexity. The architecture begins with an input layer that processes single-channel rs-fMRI images of size \( H \times W \times 1 \). Feature extraction uses three convolutional layers with filter sizes of \( 3 \times 3 \), progressively increasing in depth from 16 to 64 filters. Each convolutional layer is followed by batch normalization to stabilize the learning process and ReLU activation to introduce non-linearity. 

To progressively reduce spatial dimensions and retain essential information, max-pooling layers are introduced after each convolutional block. Max-pooling selects the highest activation in a given region, enabling the network to achieve spatial invariance while reducing computational complexity. The network employs a fully connected layer with 64 neurons and ReLU activation following feature extraction. To enhance generalization and prevent overfitting, a dropout layer with a rate of 0.3 is incorporated, stochastically deactivating neurons during training. The final classification is performed by a softmax layer, which assigns probability scores to each age group:

\begin{equation}
    P_i = \frac{e^{z_i}}{\sum_{j} e^{z_j}},
\end{equation}

where \( z_i \) represents the logit for class \( i \). The network is trained using the Adam optimizer with a learning rate of 0.0001, a mini-batch size of 32, and 100 epochs. This architecture is particularly suitable for applications where computational efficiency is a priority.

\subsubsection{DeepReg-CNN}
The DeepReg-CNN model extends the capabilities of Shallow-CNN by increasing network depth and incorporating additional regularization techniques. The architecture begins with an input layer for single-channel rs-fMRI data of size \( H \times W \times 1 \). The feature extraction process is conducted through multiple convolutional layers, where the filter sizes remain \( 3 \times 3 \), but the number of filters increases progressively (16, 32, 64, and 128). 

A key characteristic of DeepReg-CNN is the inclusion of consecutive convolutional layers within the same block, which enhances hierarchical feature learning. Unlike Shallow-CNN, where a pooling operation follows each convolutional layer, DeepReg-CNN stacks multiple convolutional layers before downsampling. This strategy enables the model to learn more abstract and complex features at deeper levels. Max-pooling layers are then applied at strategic depths to reduce spatial dimensions progressively.

Multiple dropout layers with varying rates (0.2, 0.7, 0.5, and 0.3) are introduced to address overfitting. Dropout randomly deactivates neurons during training, ensuring the network does not become overly reliant on specific features. The fully connected layers in DeepReg-CNN are more extensive than Shallow-CNN, containing 512, 128, and 64 neurons, respectively. Each fully connected layer is followed by batch normalization, which normalizes activations and accelerates convergence. The batch normalization process is mathematically defined as:

\begin{equation}
    \mu_B = \frac{1}{m} \sum_{i=1}^{m} X_i, \quad 
    \sigma_B^2 = \frac{1}{m} \sum_{i=1}^{m} (X_i - \mu_B)^2,
\end{equation}

\begin{equation}
    \hat{X}_i = \frac{X_i - \mu_B}{\sqrt{\sigma_B^2 + \epsilon}},
\end{equation}

where \( X_i \) represents the input, \( \mu_B \) and \( \sigma_B^2 \) are the batch statistics, and \( \epsilon \) is a small constant for numerical stability.

The final classification is performed using a softmax layer. Unlike Shallow-CNN, which employs a smaller capacity network, DeepReg-CNN is trained with an initial learning rate of 0.001, a mini-batch size of 64, and 50 epochs, ensuring robust feature extraction and improved generalization.

Both architectures offer distinct advantages based on computational constraints and classification performance. Shallow-CNN provides an efficient baseline to extract essential spatial features with minimal depth. Its low computational footprint makes it suitable for real-time applications and deployment on resource-limited devices. However, due to its relatively shallow depth, the model may struggle to capture highly abstract features required for complex classification tasks. In contrast, DeepReg-CNN demonstrates superior generalization performance, which is attributed to its deeper architecture and regularization strategies. Adding multiple convolutional layers allows it to learn hierarchical representations, while dropout and batch normalization ensure robust training. Empirical results indicate that DeepReg-CNN consistently outperforms Shallow-CNN in terms of classification accuracy. However, the increased complexity of DeepReg-CNN demands higher computational resources, making it less feasible for lightweight applications.

\subsection{Deep Feature Embedding with Adaptive Learning}

Deep Feature Embedding with Adaptive Learning (DFEAL) is designed to enhance EEG-based age group classification by integrating deep learning with traditional machine learning. The methodology comprises three main stages: deep feature extraction using MobileNetV2, dimensionality reduction via Principal Component Analysis (PCA) \cite{PCA}, and classification using traditional machine learning algorithms.

\subsubsection{Deep Feature Extraction using MobileNetV2}
MobileNetV2, a lightweight deep convolutional neural network optimized for computational efficiency, is employed to extract high-level features from EEG-based representations. The network is pre-trained on ImageNet and fine-tuned to learn domain-specific representations from EEG-derived feature matrices.

Given an input image $X \in \mathbb{R}^{H \times W \times 3}$, MobileNetV2 applies a sequence of depthwise separable convolutions to extract hierarchical features, represented as a feature embedding $F \in \mathbb{R}^{d}$:
\begin{equation}
    F = \text{MobileNetV2}(X; \theta),
\end{equation}
where $\theta$ represents the network parameters, and $d$ is the dimensionality of the extracted features.

\subsubsection{Dimensionality Reduction using Principal Component Analysis (PCA)}
The extracted feature set is typically high-dimensional and may contain redundant information. PCA is employed to transform the feature space into a lower-dimensional representation while preserving essential variance. Given a feature matrix $F \in \mathbb{R}^{N \times d}$, where $N$ is the number of samples, PCA computes eigenvectors of the covariance matrix:
\begin{equation}
    \Sigma = \frac{1}{N} \sum_{i=1}^{N} (F_i - \mu)(F_i - \mu)^T,
\end{equation}
where $\mu$ is the mean feature vector. The feature transformation is then performed using the top $k$ principal components:
\begin{equation}
    F' = F W_k,
\end{equation}
where $W_k \in \mathbb{R}^{d \times k}$ consists of the $k$ eigenvectors corresponding to the largest eigenvalues.

\subsection{Traditional Machine Learning Classification}
The reduced feature set is classified using three traditional machine learning models: K-Nearest Neighbors (KNN), Support Vector Machine (SVM), and Random Forest (RF). These models leverage the refined feature space for improved classification performance.

\textbf{K-Nearest Neighbors (KNN)}: The KNN classifier assigns a class label based on the majority vote among the $k$ nearest neighbors in the feature space. The decision function is defined as:
\begin{equation}
    y = \arg\max_{c} \sum_{i \in N_k} \mathbb{1}(y_i = c),
\end{equation}
where $N_k$ represents the set of $k$ nearest neighbors, and $\mathbb{1}(y_i = c)$ is an indicator function.

\textbf{Support Vector Machine (SVM)}: SVM constructs a hyperplane to separate classes in the transformed feature space by solving:
\begin{equation}
    \min_{w, b} \frac{1}{2} \|w\|^2 + C \sum_{i=1}^{N} \xi_i,
\end{equation}
subject to:
\begin{equation}
    y_i (w^T F'_i + b) \geq 1 - \xi_i, \quad \xi_i \geq 0,
\end{equation}
where $w$ and $b$ define the decision boundary, and $\xi_i$ are slack variables allowing for soft-margin classification.

\textbf{Random Forest (RF)}: RF constructs an ensemble of decision trees to enhance classification robustness. The final class prediction is obtained via majority voting among $T$ individual trees:
\begin{equation}
    y = \arg\max_c \sum_{t=1}^{T} \mathbb{1}(h_t(F') = c),
\end{equation}
where $h_t$ represents the classification output of the $t$-th decision tree.

Empirical evaluations demonstrated that the DFEAL framework significantly improves classification accuracy compared to standalone deep learning models. By leveraging deep feature extraction, dimensionality reduction, and adaptive learning through traditional classifiers, DFEAL enhances generalization while mitigating overfitting. 
\subsection{Brain Network Visualization based on SBPS}

To visualize the average brain networks of different age cohorts, we categorized participants into four groups: Younger Adults (Y: 7-19 years), Middle Young (MY: 20-34 years), Middle Late (ML: 35-53 years), and Elderly (E: 54-89 years) based on age clusters. BrainNet Viewer, a graph-theoretical network visualization toolbox, was used to represent human connectomes as ball-and-stick models \cite{xia2013brainnet}. The methodology involved several steps. Firstly, the raw rs-fMRI time-series data for each participant were extracted based on the Dosenbach brain anatomical parcellation. Next, seed-based phase synchronization between different ROI time-series data was computed. A threshold analysis was performed based on a graphical approach (number of edges versus threshold value), and an absolute threshold of 0.2 was chosen, as previously used in studies for binarizing connectivity matrices \cite{azarmi2019granger}. After thresholding, the binarized phase-synchronized matrices for different subjects within each age cohort were averaged, and an edge file was generated for visualization. Each age group’s mean SBPS matrix was analyzed to visualize connectivity patterns, ensuring a clear understanding of brain network changes across developmental stages. The resulting visualization effectively highlights how functional connectivity evolves from young adulthood to late aging.

\subsection{Clustering of each Brain Developmental Stages}

We performed clustering of brain developmental stages based on brain connectivity features extracted for each age cohort, utilizing Z-transformed Pearson correlation matrices, seed-based phase-synchronized (SBPS) matrices, and their combined features to assess clustering improvements. Previous studies have primarily used Pearson correlation for age group clustering \cite{singh2023reorganization}, but comparing it with SBPS provides additional insights into connectivity patterns and improved visualization of age-related differences. We systematically processed data from four age-based clusters—Young, Middle Young, Middle-Late, and Elderly—across two feature types. For each cluster, we extracted matrices, handled inconsistencies, computed mean matrices, and concatenated data types for each feature. The final combined matrices were analyzed using t-SNE for visualization \cite{van2008visualizing}, ensuring reproducibility through fixed random seeds. The results help evaluate how clustering performance improves with different feature combinations. By comparing the clustering performance of Pearson correlation, SBPS, and their combined features, we assessed how feature selection impacts the clarity of age group separation. This analysis highlights the advantages of incorporating multiple connectivity measures for robust clustering.

\subsection{Study of Brain Functional Network Measures with Age}

In this study, we investigate key network measures—clustering coefficient, participation coefficient, global efficiency, integration, and modularity—derived from resting-state fMRI correlation matrices. These metrics provide insights into network segregation, integration, and efficiency, helping to understand age-related cognitive changes. The fMRI data for 1096 subjects were preprocessed, and correlation matrices were derived from the region-wise time series data. Pearson’s correlation was used to compute the correlation coefficients between 160 ROIs (Dosenbach atlas), forming subject-specific functional connectivity matrices. The Pearson correlation coefficient between two ROIs \(i\) and \(j\) was computed as:

\begin{equation}
    r_{ij} = \frac{\sum_{t} (X_i(t) - \bar{X}_i)(X_j(t) - \bar{X}_j)}
    {\sqrt{\sum_{t} (X_i(t) - \bar{X}_i)^2} \sqrt{\sum_{t} (X_j(t) - \bar{X}_j)^2}}
\end{equation}

where \(X_i(t)\) and \(X_j(t)\) represent the time series of ROIs \(i\) and \(j\), respectively.

To construct functional brain networks, these correlation matrices were thresholded at values ranging from 0.1 to 0.5 in increments of 0.05, setting connections below the threshold to zero to ensure sparse network representations.

After that following Graph-Theoretic Measures were computed:

\subsubsection{Clustering Coefficient}

The clustering coefficient quantifies the likelihood of nodes forming local clusters and is computed as \cite{watts1998collective}:

\begin{equation}
    C_i = \frac{2T_i}{k_i (k_i - 1)}
\end{equation}

where \(T_i\) represents the number of triangles including node \(i\), and \(k_i\) is the node degree.

\subsubsection{Participation Coefficient}

The participation coefficient reflects how evenly a node distributes its connections across different communities \cite{rubinov2010complex}:

\begin{equation}
    P_i = 1 - \sum_{m} \left(\frac{k_{i,m}}{k_i}\right)^2
\end{equation}

where \(k_{i,m}\) is the strength of connections between node \(i\) and community \(m\), and \(k_i\) is the total degree of node \(i\).

\subsubsection{Global Efficiency}

Global efficiency measures network-wide information transfer and is given by \cite{bullmore2009complex}:

\begin{equation}
    E_{glob} = \frac{1}{N(N-1)} \sum_{i \neq j} \frac{1}{d_{ij}}
\end{equation}

where \(d_{ij}\) represents the shortest path length between nodes \(i\) and \(j\).

\subsubsection{Integration}

The integration metric was computed from the determinant of the correlation matrix \cite{sporns2013network}:

\begin{equation}
    I = \log \det (C)
\end{equation}

where \(C\) is the positive semi-definite correlation matrix obtained after eigenvalue decomposition.

\subsubsection{Modularity}

Modularity quantifies the degree of modular structure within the network \cite{newman2006modularity}:

\begin{equation}
    Q = \frac{1}{2m} \sum_{i,j} \left[A_{ij} - \frac{k_i k_j}{2m} \right] \delta(c_i, c_j)
\end{equation}

where \(A_{ij}\) represents the adjacency matrix, \(k_i\) and \(k_j\) are the degrees of nodes \(i\) and \(j\), \(m\) is the total number of edges, and \(\delta(c_i, c_j)\) is 1 if nodes \(i\) and \(j\) belong to the same community and 0 otherwise. Community detection was performed using k-means clustering with two communities.

To analyze the variation of these network measures with age, we extracted subject ages from a predefined dataset (\texttt{AGEMATRIX}). Outliers were removed using Z-score thresholding. The relationship between network metrics and age was analyzed using quadratic polynomial fitting \cite{zhan2015comparison}:

\begin{equation}
    y = a x^2 + b x + c
\end{equation}

where \( y \) represents each network metric (clustering coefficient, participation coefficient, global efficiency, integration, modularity) and \( x \) represents age.

Scatter plots were generated with polynomial trend lines to visualize the association between age and each network measure, highlighting potential nonlinear trends.

\section{Results and Discussions}
\subsection{Experimental Setup}
The experiments were conducted in MATLAB, utilizing toolboxes for image processing, machine learning, deep network design, and Brainstorm. The implementation ran on a high-performance system featuring an Intel i7-13650HX processor, 64 GB RAM, and an NVIDIA Quadro P5000 GPU with 16 GB of dedicated memory. This robust computational setup facilitated efficient data processing, extensive model evaluation, and reliable performance across all experimental analyses. 

\subsection{Evaluation of Various CNNs}
To assess the efficacy of different CNNs in fMRI-based age group classification, we conducted a comprehensive evaluation using MobileNetV2, Shallow-CNN, and DeepReg-CNN across multiple feature representations, including SBPS, Z-Transformed Features (ZTF), Raw fMRI, and FMCR data. The classification performance of each model was quantified in terms of training and validation accuracy, as summarized in Table \ref{tab:CNNsEvaluation}.

\begin{table}[h]
    \centering
    \caption{CNN-Based Approaches for fMRI-Based Age Group Classification}
    \begin{adjustbox}{max width=\textwidth}
    \begin{tabular}{l l c c}
        \toprule
        \textbf{Model} & \textbf{Approach} & \textbf{Train Acc (\%)} & \textbf{Val Acc (\%)}\\
        \midrule
        MobileNetV2  & SBPS & 95.56  & 60.23 \\
        Shallow-CNN & SBPS  & 99.89  & 52.80 \\
        DeepReg-CNN & SBPS  & 93.88  & 62.80 \\
        MobileNetV2  & ZTF & 77.98 & 60.23\\
        Shallow-CNN & ZTF & 60.00  & 58.23\\
        DeepReg-CNN & ZTF & 69.09  & 63.33\\
        MobileNetV2  & Raw fMRI & 97.36  & 82.15\\
        Shallow-CNN & Raw fMRI & 68.23  & 67.80 \\
        DeepReg-CNN & Raw fMRI & 95.50  & 76.89 \\
        MobileNetV2  & FMCR & 99.23  & 85.88 \\
        Shallow-CNN & FMCR & 99.12  & 79.55\\
        DeepReg-CNN & FMCR & 96.50  & 84.32 \\
        \bottomrule
    \end{tabular}
    \end{adjustbox}
    \label{tab:CNNsEvaluation}
\end{table}

The results indicate a substantial variation in model performance depending on both the network architecture and the type of input feature representation. In general, models trained on raw and FMCR data achieved superior validation accuracy compared to those utilizing SBPS or ZTF, suggesting that preserving spatial and temporal integrity in the input representation enhances classification robustness. 

MobileNetV2 consistently demonstrated superior generalization capability across all feature representations. Notably, when trained on FMCR data, MobileNetV2 achieved a validation accuracy of \textbf{85.88\%}, outperforming both Shallow-CNN (79.55\%) and DeepReg-CNN (84.32\%). Similarly, for Raw fMRI data, MobileNetV2 attained an accuracy of \textbf{82.15\%}, surpassing DeepReg-CNN (76.89\%) and Shallow-CNN (67.80\%). This highlights the effectiveness of MobileNetV2 in extracting high-level discriminative features, particularly when trained on richer, augmented datasets.
\begin{table*}[t!]
    \centering
    \caption{MobileNet + PCA + Random Forest with K-Fold Cross-Validation (Three-Channel Input: 160x160x3)}
    \begin{adjustbox}{max width=\textwidth}
    \begin{tabular}{c c c c c c c c c c c c}
        \toprule
        \textbf{NoT} & \textbf{F1} & \textbf{F2} & \textbf{F3} & \textbf{F4} & \textbf{F5} & \textbf{F6} & \textbf{F7} & \textbf{F8} & \textbf{F9} & \textbf{F10} & \textbf{Avg Acc (\%)} \\
        \midrule
        50  & 0.8872  & 0.8906  & 0.8845  & 0.9027  & 0.8875  & 0.9058  & 0.9179  & 0.8723  & 0.9058  & 0.8719  & 0.8926 \\
        100 & 0.9024  & 0.8936  & 0.8815  & 0.8754  & 0.8815  & 0.9179  & 0.8693  & 0.8997  & 0.9210  & 0.8750  & 0.8917 \\
        150 & 0.8719  & 0.8967  & 0.8997  & 0.8723  & 0.9058  & 0.8875  & 0.9179  & 0.9149  & 0.8632  & 0.8963  & 0.8926 \\
        200 & 0.8963  & 0.8784  & 0.9119  & 0.8967  & 0.8723  & 0.8936  & 0.8875  & 0.8967  & 0.8875  & 0.8750  & 0.8896 \\
        250 & 0.8689  & 0.8967  & 0.8906  & 0.9210  & 0.8906  & 0.8784  & 0.8784  & 0.9179  & 0.8693  & 0.9299  & 0.8942 \\
        300 & 0.9055  & 0.9271  & 0.9179  & 0.9119  & 0.8875  & 0.8967  & 0.8845  & 0.8997  & 0.8663  & 0.9024  & 0.8999 \\
        400 & 0.9085  & 0.8967  & 0.8875  & 0.8784  & 0.9210  & 0.8815  & 0.8967  & 0.8906  & 0.9058  & 0.8598  & 0.8926 \\
        \bottomrule
    \end{tabular}
    \end{adjustbox}
    \label{tab:DEFAL}
\end{table*}
Conversely, for SBPS-based features, all models exhibited a significant performance drop, with the best validation accuracy reaching only 62.80\% (DeepReg-CNN). This suggests that SBPS alone may not be sufficiently expressive for capturing age-related resting state fMRI data patterns. A similar trend is observed for ZTF, where DeepReg-CNN achieved the highest validation accuracy of 63.33\%, emphasizing that statistical transformations such as ZTF, while useful, may require complementary features for optimal classification.

The results reveal that Shallow-CNN, despite achieving high training accuracy, suffers from poor generalization, as evidenced by the drastic performance gap between training and validation accuracies. For instance, when trained on SBPS features, Shallow-CNN reached an overfitted \textbf{99.89\%} training accuracy but yielded only \textbf{52.80\%} validation accuracy. This suggests that its limited depth may prevent effective feature learning, particularly in high-dimensional fMRI data.

DeepReg-CNN, designed to incorporate additional regularization and deeper feature extraction, exhibited a better balance between training and validation performance compared to Shallow-CNN. However, it was still outperformed by MobileNetV2 in most cases, suggesting that deeper CNNs with pre-trained feature extractors can further enhance classification robustness.

The experimental findings underscore that MobileNetV2, coupled with FMCR data, provides the most reliable classification performance, achieving the highest validation accuracy (85.88\%). The results also emphasize that raw fMRI signals preserve essential spatial-temporal characteristics crucial for accurate classification, whereas SBPS and ZTF alone may not be sufficient. These insights motivate the need for feature fusion techniques to further improve classification performance, as explored in the subsequent sections.

\subsection{Evaluation of DFEAL}
The proposed DFEAL approach was implemented to enhance classification performance further, integrating deep feature extraction, dimensionality reduction, and traditional machine learning classifiers. This method involves extracting bottleneck features using MobileNetV2, applying PCA for dimensionality reduction, and employing Random Forest (RF) for classification. The evaluation was performed using a 10-fold cross-validation strategy, and the classification performance was analyzed by varying the number of trees (\textbf{NoT}) in the RF classifier.

The classification results for different NoT values are presented in Table \ref{tab:DEFAL}, where the accuracy across different folds (\textbf{F1}–\textbf{F10}) is reported along with the average accuracy. The FMCR approach was utilized as feature engineering. This fusion allowed the model to capture complementary information from distinct feature spaces, improving classification robustness.

The results indicate that the optimal performance is achieved with NoT = 50, yielding an average accuracy of \textbf{89.33\%} across all folds. The model performance remains consistent for NoT = 100 and NoT = 150, achieving average accuracies of 89.17\% and 89.26\%, respectively. While minor fluctuations in fold-wise accuracy exist, the overall performance remains stable across different RF configurations.

\begin{figure}[t!]
    \centering
    \includegraphics[width=1\linewidth, trim=100 220 120 245,clip]{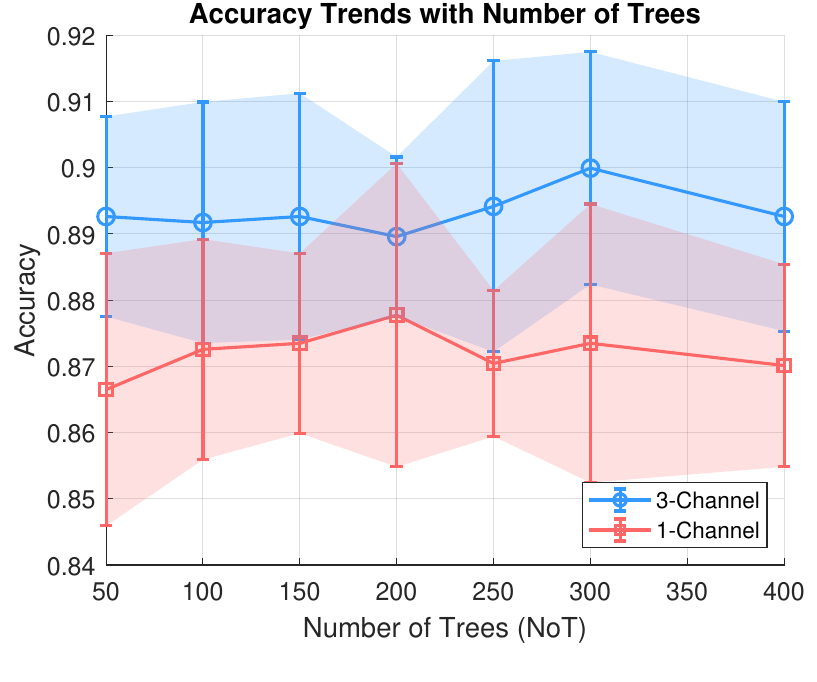}
    \caption{DFEAL Performance across NoT}
    \label{fig:Performance of DFEAL across trees}
\end{figure}

\begin{figure*}[t]
\includegraphics[width=\linewidth, keepaspectratio = True]{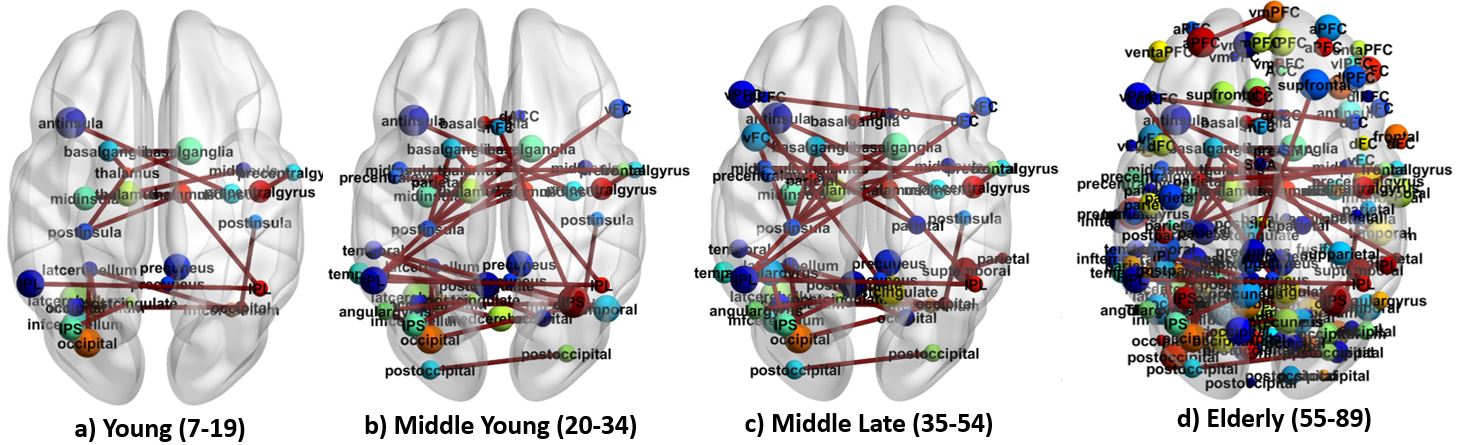}
\caption{(a), (b), (c), and (d) represent brain network connectivity based on Phase Synchronization for 160 dos ROIs of Young (Y), Middle Young (MY), (Middle late) ML, and Elder (E) stages respectively. With increasing age anterior brain nodes are showing more connectivity than young age.}
\label{figure:figure ps}
\end{figure*}

The observed improvements in classification accuracy can be attributed to integrating multiple feature representations. Unlike single-modality input approaches, the three-channel representation leverages both spatial and statistical properties of fMRI signals, enhancing the model’s ability to discern age-related patterns. The PCA transformation effectively reduces redundancy in deep features, ensuring that the most informative components are retained for classification.

Compared to direct CNN-based classifiers (as discussed in the previous subsection), DFEAL exhibits a 3\% improvement in classification accuracy, further reinforcing the effectiveness of hybrid approaches that combine deep learning with traditional machine learning techniques. The MobileNetV2 model alone, when trained end-to-end, achieved a maximum validation accuracy of 85.88\%, whereas the DFEAL pipeline, integrating PCA and RF, boosted the accuracy to 89.33\%.

Increasing NoT beyond 50 does not result in significant improvements, suggesting that a higher number of decision trees may introduce redundancy without providing additional discriminative power. The slight variations in accuracy across different folds indicate that RF remains a robust and stable classifier, effectively utilizing the deep feature representations extracted from MobileNetV2.

The experimental findings highlight that DFEAL effectively enhances classification performance by leveraging deep feature extraction, dimensionality reduction, and ensemble learning. Combining MobileNetV2 bottleneck features, PCA-based dimensionality reduction, and RF classification balances computational efficiency and predictive accuracy. This approach demonstrates the potential of hybrid learning paradigms in resting state fMRI-based age group classification, outperforming standalone deep learning models.

\begin{table*}[t!]
    \centering
    \caption{MobileNet + PCA + Random Forest with K-Fold Cross-Validation (Single-Channel Input: 160x160x1)}
    \begin{adjustbox}{max width=\textwidth}
    \begin{tabular}{c c c c c c c c c c c c}
        \toprule
        \textbf{NoT} & \textbf{F1} & \textbf{F2} & \textbf{F3} & \textbf{F4} & \textbf{F5} & \textbf{F6} & \textbf{F7} & \textbf{F8} & \textbf{F9} & \textbf{F10} & \textbf{Avg Acc (\%)} \\
        \midrule
        50  & 0.8781  & 0.8480  & 0.8571  & 0.8663  & 0.8784  & 0.9088  & 0.8359  & 0.8511  & 0.8632  & 0.8781  & 0.8665 \\
        100 & 0.8933  & 0.8815  & 0.8936  & 0.8663  & 0.8632  & 0.8754  & 0.8754  & 0.8754  & 0.8359  & 0.8659  & 0.8726 \\
        150 & 0.8842  & 0.8663  & 0.8936  & 0.8571  & 0.8815  & 0.8632  & 0.8511  & 0.8784  & 0.8845  & 0.8750  & 0.8735 \\
        200 & 0.8750  & 0.8298  & 0.9119  & 0.9088  & 0.8663  & 0.8875  & 0.8754  & 0.8723  & 0.8723  & 0.8781  & 0.8777 \\
        250 & 0.8811  & 0.8815  & 0.8845  & 0.8663  & 0.8754  & 0.8571  & 0.8723  & 0.8511  & 0.8723  & 0.8628  & 0.8704 \\
        300 & 0.8750  & 0.8693  & 0.8663  & 0.8906  & 0.8602  & 0.8328  & 0.9058  & 0.8906  & 0.8571  & 0.8872  & 0.8735 \\
        400 & 0.8567  & 0.8632  & 0.8815  & 0.8571  & 0.8875  & 0.8906  & 0.8875  & 0.8480  & 0.8663  & 0.8628  & 0.8701 \\
        \bottomrule
    \end{tabular}
    \end{adjustbox}
    \label{tab: Singlechannel}
\end{table*}

\subsection{Ablation Study and Comparative Analysis}

To gain deeper insights into the effectiveness of the proposed DFEAL approach, we conducted an ablation study by evaluating the impact of different input configurations on classification performance. Additionally, we performed a comparative analysis by benchmarking various classifiers, including KNN, SVM, and RF, to determine the optimal choice for resting-state time series fMRI-based age group classification.

To assess the contribution of feature fusion in DFEAL, we compared the classification performance of the Random Forest classifier using single-channel input (160×160×1) versus three-channel input (160×160×3). Table \ref{tab: Singlechannel} presents the classification results for single-channel input, while Table \ref{tab:CNNsEvaluation} (discussed in the previous subsection) details the results for the three-channel configuration.
\begin{table*}[t!]
    \centering
    \caption{Comparative Performance of Different Classifiers}
    \begin{adjustbox}{max width=\textwidth}
    \begin{tabular}{l c c c c c c c c c c c}
        \toprule
        \textbf{Classifier} & \textbf{F1} & \textbf{F2} & \textbf{F3} & \textbf{F4} & \textbf{F5} & \textbf{F6} & \textbf{F7} & \textbf{F8} & \textbf{F9} & \textbf{F10} & \textbf{Avg Acc (\%)} \\
        \midrule
        KNN  & 0.7317 & 0.7264 & 0.7690 & 0.8055 & 0.7629 & 0.7903 & 0.7812 & 0.7477 & 0.7720 & 0.7256 & \textbf{0.7612} \\
        SVM  & 0.7134 & 0.7508 & 0.7325 & 0.7508 & 0.7386 & 0.7568 & 0.7599 & 0.7477 & 0.7204 & 0.7683 & 0.7439 \\
        Random Forest & 0.9055 & 0.9271 & 0.9179 & 0.9119 & 0.8875 & 0.8967 & 0.8845 & 0.8997 & 0.8663 & 0.9024 & \textbf{0.8999} \\
        \bottomrule
    \end{tabular}
    \end{adjustbox}
    \label{tab:classifiers}
\end{table*}

The results indicate that the three-channel representation significantly outperforms the single-channel input, achieving an average accuracy of 89.33\% compared to 87.21\% with single-channel input. This improvement underscores the advantage of incorporating multiple feature matrices (raw fMRI, ZTF, and phase synchronization) to capture complementary aspects of fMRI dynamics. The fusion of diverse representations enhances feature richness, enabling the model to learn more discriminative patterns associated with different age groups.

Furthermore, the single-channel approach exhibits greater variability across different folds, with certain folds showing a notable drop in accuracy. This suggests that relying on a single feature representation may lead to inconsistent generalization across various subsets of the dataset.

To evaluate the effectiveness of various machine learning classifiers in the DFEAL framework, we tested KNN, SVM, and RF on the extracted deep features. Table \ref{tab:classifiers} summarizes the classification performance across different folds.

The results demonstrate that Random Forest consistently outperforms KNN and SVM, achieving an average accuracy of 89.99\%, which is 13.87\% higher than KNN (76.12\%) and 15.60\% higher than SVM (74.39\%). This performance advantage can be attributed to the ensemble nature of RF, which effectively captures complex decision boundaries by aggregating multiple decision trees, thereby reducing overfitting and improving robustness.

In contrast, KNN and SVM exhibit lower performance, with KNN performing slightly better than SVM. The relatively weaker performance of these classifiers suggests that deep feature representations extracted from MobileNetV2 are inherently nonlinear, and traditional linear classifiers (such as SVM) struggle to separate the age groups in the transformed feature space effectively.

\begin{figure*}[h]
\centering
\begin{subfigure}[h]{0.49\linewidth}
\centering
\includegraphics[width =\linewidth , keepaspectratio = True]{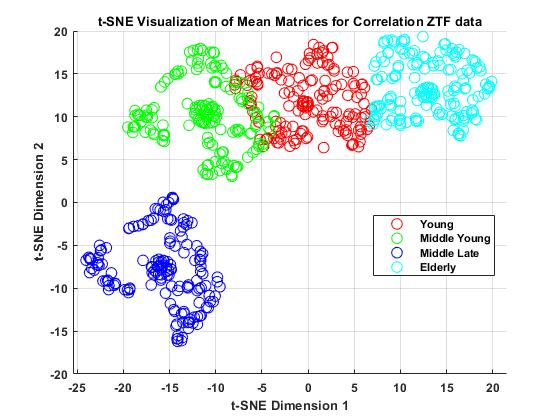}
\label{figure:CZ}
\caption{Clustering of BDS based on ZTF Pearson correlation data.}
\end{subfigure}
\begin{subfigure}[h]{0.49\linewidth}
\centering
\includegraphics[width =\linewidth , keepaspectratio = True]{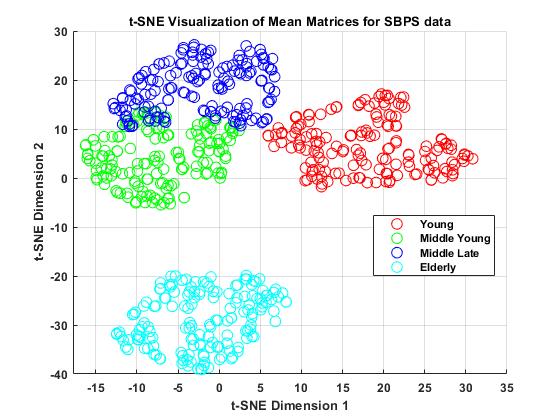}
\label{figure:sbps}
\caption{Clustering of BDS based on SBPS data. } 
\end{subfigure}
\begin{subfigure}[h]{0.49\linewidth}
\centering
\includegraphics[width =\linewidth , keepaspectratio = True]{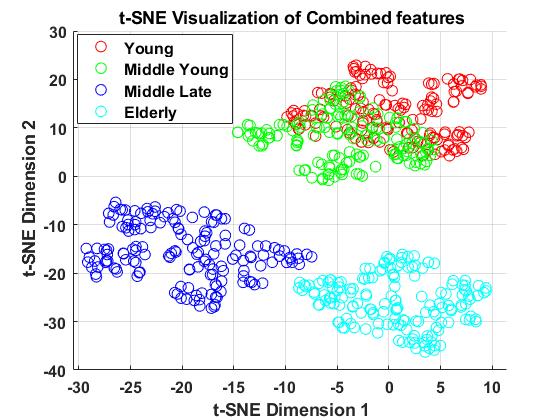}
\label{figure:combined}
\caption{Clustering of BDS based on combined ZTF Pearson correlation and SBPS. } 
\end{subfigure}
\label{figure: clustering}
\caption{Clustering and Plotting of four BDS based on t-SNE clustering.}
\end{figure*}

\begin{figure*}[t]
\includegraphics[width=\linewidth, , keepaspectratio = True]{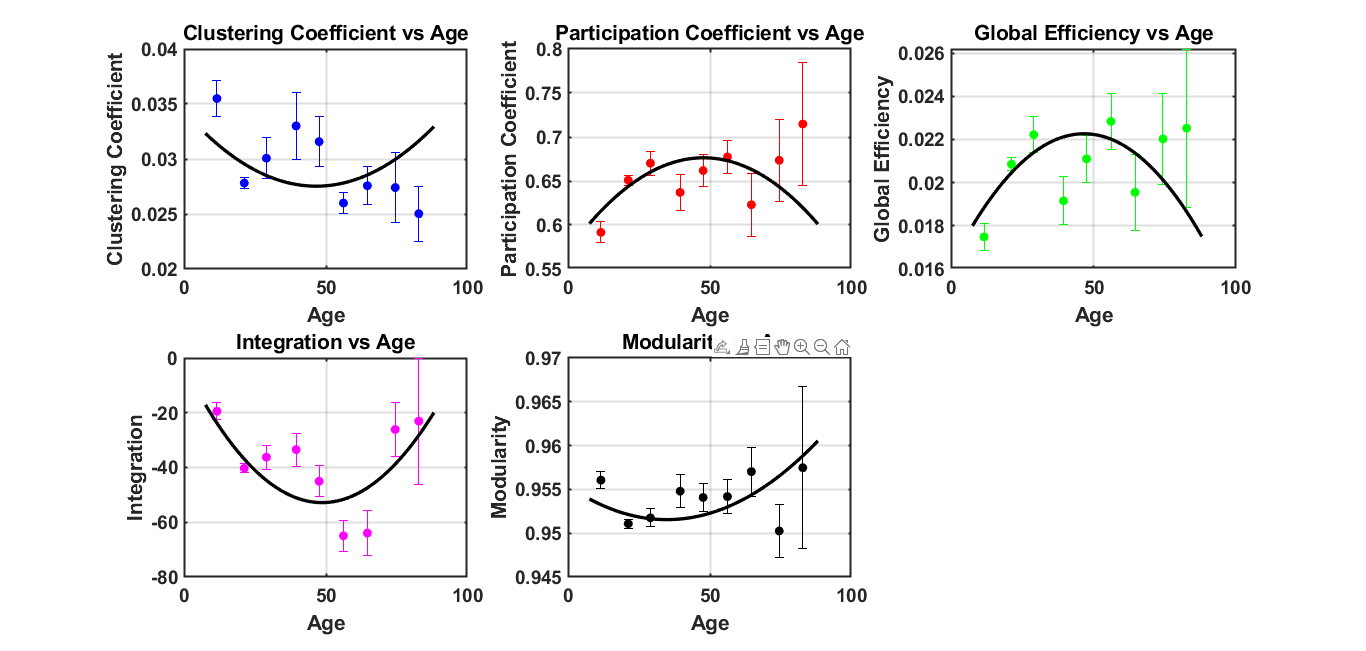}
\caption{Variation of correaltion based functional connectivity metrics with increasing age.}
\label{figure:Age trend analysis}
\end{figure*}

\subsection{Phase Synchronization Across Age Groups}
Phase synchronization, a key indicator of functional connectivity, undergoes notable shifts with aging, reflecting changes in neural communication and network organization. Previous studies using phase-based connectivity measures, such as phase-locking value (PLV) and weighted phase-lag index (wPLI), have reported a decline in long-range synchronization and an increase in local connectivity with aging, suggesting a transition from a globally integrated to a more fragmented network structure \cite{stam2007phase, vecchio2014human}. These changes have been linked to age-related alterations in neural efficiency, reduced network integration, and potential compensatory mechanisms aimed at maintaining cognitive function \cite{gaal2010age}.

Consistent with these findings, our results, as shown in Figure 3, reveal distinct phase synchronization patterns across different age groups at a stronger synchronization threshold (0.3). In young adults, synchronization is primarily observed in subcortical, occipital, and cerebellar nodes, indicating a reliance on sensory and subcortical processing hubs. This pattern suggests efficient information transfer through well-established neural pathways.

In the middle-young group, phase synchronization extends to temporal, central, occipital, post-occipital, and cerebellar nodes, suggesting increased recruitment of cortical regions. This shift may reflect the early integration of associative and executive processing networks, supporting the progressive adaptation of brain function with age.

As individuals transition to middle-late adulthood, synchronization further involves a few frontal nodes alongside temporal, central, occipital, post-occipital, and cerebellar regions. This pattern may indicate the onset of compensatory network reorganization due to age-related declines in long-range connectivity, as previously reported in resting-state fMRI studies \cite{sala2015reorganization}.

In the elderly group, synchronization is most widespread, encompassing prefrontal, frontal, temporal, central, occipital, post-occipital, and cerebellar nodes. This broader distribution aligns with studies reporting increased local connectivity and potential compensatory mechanisms to preserve cognitive function in aging populations \cite{bullmore2009complex,meunier2009age}. The greater involvement of frontal regions supports the notion of frontal over-recruitment, often associated with compensatory neural responses during aging \cite{cabeza2002aging}.

These findings suggest a progressive shift from a sensory-driven network in youth to a more distributed synchronization pattern with age, potentially compensating for declining global efficiency in functional connectivity. The observed age-related reorganization of phase synchronization highlights the dynamic nature of brain networks and their adaptation to neurobiological changes across the lifespan.

\subsection{Clustering of BDS using t-SNE}
Understanding the developmental trajectory of brain networks is crucial for studying age-related changes in functional connectivity. Previous studies have demonstrated that brain connectivity features derived from resting-state fMRI can effectively capture age-related variations in network organization \cite{betzel2014changes, fair2008maturing}. In particular, seed-based functional connectivity and phase synchronization measures have been widely used to assess developmental changes in neural communication \cite{gao2017functional,meunier2009age}.

This study applied t-distributed Stochastic Neighbor Embedding (t-SNE) to different brain connectivity features to cluster and visualize brain developmental stages (BDS). Specifically, we used (i) Pearson correlation z-transformed (ZTF) connectivity matrices, (ii) seed-based phase synchronization (SBPS) data for 160 Dosenbach ROIs, and (iii) a combined feature set of both connectivity measures as shown in figure 4 (a), 4(b) and 4(c) respectively. t-SNE successfully distinguished different BDS, with SBPS-based features yielding the most well-defined clusters, followed by Correlation ZTF and the combined feature set.

These findings align with previous research demonstrating that phase-based connectivity measures, such as SBPS, are more sensitive to age-related changes in functional integration and segregation \cite{vecchio2014human, sala2015reorganization}. The observed clustering pattern suggests that brain connectivity features can effectively distinguish various age groups and could serve as reliable markers for predicting developmental trajectories. The superior clustering performance of SBPS further supports its relevance in capturing dynamic neural interactions and age-related reorganization of functional networks.

\subsection{Age-Related Variation in Functional Connectivity Metrics}

This study examined resting-state functional MRI (rs-fMRI) data from 1,096 subjects to explore how key network metrics—clustering coefficient, participation coefficient, global efficiency, integration, and modularity—evolve with age, as shown in Figure 5. These measures provide crucial insights into network segregation, integration, and efficiency, which play a fundamental role in cognitive aging. We employed scatter plots with quadratic polynomial fits to capture nonlinear trends across age groups.

 \textbf{a) Clustering Coefficient}
The clustering coefficient, which represents local segregation within brain networks, followed a U-shaped trajectory across the lifespan. It declined from young adulthood (18–35 years), reaching its lowest point around midlife (~50 years), and then increased in older age (~80 years), as shown in Figure 5. This pattern aligns with previous findings that suggest a decline in local connectivity efficiency during midlife due to cortical thinning and synaptic pruning, followed by a compensatory increase in later years \cite{song2014age, betzel2014changes}.

\textbf{b) Participation Coefficient }
The participation coefficient, a measure of cross-modular integration, exhibited an inverted U-shaped trend, as observed in Figure 5. It increased during young adulthood, peaked around midlife (~50 years), and subsequently declined with aging. This pattern suggests that middle-aged adults achieve the most efficient network integration, whereas aging leads to reduced intermodular connectivity \cite{betzel2019structural, chan2014decreased}. The decline in older adults implies that brain regions become more functionally specialized, forming fewer connections across modules, a trend consistent with prior research on reduced network integration in aging populations.

\textbf{c) Global Efficiency}
Global efficiency, indicative of the brain’s ability to facilitate information transfer across the network, also followed an inverted U-shaped pattern. It peaked around midlife (~40–50 years) before declining in older age as observed in figure 5. The observed reduction in later years likely results from white matter degeneration and diminished long-range connectivity \cite{achard2007efficiency}. This trend is consistent with previous findings that associate aging with decreased global efficiency, which may contribute to cognitive slowing and processing deficits in older adults \cite{gu2022overlapping}.

\textbf{d) Integration}
Network integration, quantified using the determinant of the correlation matrix, showed a distinct U-shaped pattern. It was highest in early adulthood, declined sharply around midlife (~50 years), and then partially recovered in older age, as shown in Figure 5. This pattern aligns with prior studies indicating that middle-aged individuals experience transient functional network disintegration due to structural and synaptic reorganization \cite{grady2012cognitive}. The reduction in integration during midlife may underlie declines in cognitive processes such as memory and executive function.

\textbf{e) Modularity}
Network modularity, which reflects the degree of network segregation into distinct communities, displayed a weak U-shaped trajectory, as observed in Figure 5. A slight reduction in midlife (~50 years) was followed by an increase in old age. This suggests that middle adulthood may represent a transitional phase of reduced modularity, whereas aging is characterized by a shift toward a more segregated network architecture \cite{meunier2009age}. The increased modularity in older adults may act as a compensatory mechanism to preserve cognitive function despite widespread neural declines.

Our results suggest that integration-based metrics, including the participation coefficient and global efficiency, follow an inverted U-shaped trajectory, peaking in midlife before declining in older age. In contrast, segregation-related metrics such as clustering coefficient and modularity exhibit a U-shaped trend, with a midlife dip followed by recovery in later years. These findings align with the notion that brain networks transition from a balanced integration-segregation regime in early adulthood to midlife declines, followed by compensatory reorganization in aging \cite{betzel2014changes}.

These age-related shifts in network topology underscore the dynamic nature of brain connectivity across the lifespan. The initial increase in the clustering coefficient followed by a decline, along with reductions in participation coefficient and global efficiency, suggest a transition from an integrated to a more segregated network organization with aging. Such changes may underlie cognitive aging phenomena such as slowed processing speed and reduced cognitive flexibility. Understanding these patterns is essential for developing targeted interventions to mitigate cognitive decline in aging populations.

\section{Conclusion}

This study explored age-related changes in functional brain connectivity using phase synchronization, clustering techniques, and Pearson correlation-based graph-theoretic metrics. Our findings reveal a progressive shift in phase synchronization patterns from sensory-driven networks in youth to more distributed and compensatory connectivity in aging, aligning with previous reports of altered functional network organization across the lifespan.

Through t-SNE clustering, we demonstrated that seed-based phase synchronization (SBPS) features offer superior differentiation of brain developmental stages compared to correlation-based connectivity measures. This highlights SBPS as a potential biomarker for tracking age-related changes in neural communication.

To further enhance classification performance, we implemented deep learning and machine learning approaches, including MobileNet, Random Forest, and Principal Component Analysis (PCA)-based feature reduction. The classification of brain developmental stages (BDS) using combined connectivity features showed improved accuracy with these models, with MobileNet outperforming traditional methods. This suggests that deep learning approaches effectively capture complex, nonlinear relationships in functional connectivity, making them valuable for age-related brain network analysis.

Analysis of network topology metrics further supports the notion of dynamic shifts in brain organization with aging. Integration-based measures, such as participation coefficient and global efficiency, followed an inverted U-shaped trajectory, peaking around midlife before declining. In contrast, segregation-related measures, such as clustering coefficient and modularity, exhibited a U-shaped pattern, indicating a transition from balanced network integration in early adulthood to midlife declines, followed by compensatory reorganization in aging.

However, this study has several limitations. First, only a subset of functional connectivity features was used for age group classification, potentially overlooking additional network characteristics relevant to aging. Second, the analysis was limited to resting-state fMRI data, which does not capture task-specific neural dynamics. Third, the study focused solely on functional connectivity without incorporating structural connectivity or multimodal imaging data, which could provide a more comprehensive understanding of age-related brain network changes.

Despite these limitations, our findings underscore the adaptive nature of brain networks across different age groups. The observed alterations in connectivity patterns suggest that aging involves both functional reorganization and compensatory mechanisms to maintain cognitive function. The use of deep learning for classification further highlights the potential of advanced computational approaches in understanding brain aging. Future studies should explore multimodal datasets, incorporate additional connectivity features, and investigate task-based fMRI to gain a more holistic view of brain network alterations across the lifespan.

 \section{Disclosure}
The authors have declared no conflict of interest related to this study.

 \section{Acknowledgement}
 This research work is supported by the Neurocomputing Laboratory and Multichannel Signal Processing Laboratory (MSP Lab) at the Indian Institute of Technology Delhi (IIT Delhi). Data were provided [in part] by the Human Connectome Project, WU-Minn Consortium (Principal Investigators: David Van Essen and Kamil Ugurbil; 1U54MH091657) funded by the 16 NIH Institutes and Centers that support the NIH Blueprint for Neuroscience Research; and by the McDonnell Center for Systems Neuroscience at Washington University.

\bibliographystyle{unsrt}
\bibliography{refs}

\vspace{15mm}

\end{document}